\documentclass[useAMS,usenatbib]{mn2e}

\usepackage{epsfig}
\usepackage{graphics}
\usepackage[usenames]{color}
\usepackage{endnotes}

\usepackage{amssymb}
\usepackage{amsmath}
\usepackage{times}
\usepackage{subfigure}
\newcommand{\be}{\begin{equation}}
\newcommand{\ee}{\end{equation}}
\newcommand{\bdm}{\begin{displaymath}}
\newcommand{\edm}{\end{displaymath}}
\newcommand{\bea}{\begin{eqnarray}}
\newcommand{\eea}{\end{eqnarray}}

\newcommand{\msun}{M_\odot}
\def\lsim{\lower.5ex\hbox{$\; \buildrel < \over \sim \;$}}

\title[BH eccentricity in circumbinary discs]
{Limiting eccentricity of sub-parsec massive black hole binaries surrounded by
  self-gravitating gas discs}

\author[C. R\"odig, M. Dotti, A. Sesana, J.Cuadra, M.Colpi ]
       {C. R\"odig$^{1}$\thanks{E-mail: croedig@aei.mpg.de}, M. Dotti$^{2,3}$, A. Sesana$^{1}$, J.Cuadra$^{2,4}$, M.Colpi$^3$
\\
$^{1}$ Max-Planck-Institut f{\"u}r Gravitationsphysik, Albert Einstein
Institut, Golm, Germany \\
$^{2} $ Max-Planck-Institut f\"{u}r Astrophysik, Karl-Schwarzschild-Str.\ 1, D-85748 Garching, Germany\\
$^3$ Universit\`a di Milano Bicocca, Dipartimento di Fisica
G. Occhialini, Piazza della Scienza 3, I-20126, Milano, Italy\\
$^4$ Departamento de Astronom\'ia y Astrof\'isica, Pontificia Universidad Cat\'olica de Chile, Santiago, Chile\\
}

\begin{document}

\date{}

\pagerange{\pageref{firstpage}--\pageref{lastpage}} \pubyear{2010}

\maketitle

\label{firstpage}

\begin{abstract}
We study the dynamics of supermassive black hole binaries embedded in
circumbinary gaseous discs, with the SPH code {\sc Gadget-2}.  The
sub-parsec binary (of total mass $M$ and mass ratio $q=1/3$) has
excavated a gap and transfers its angular momentum to the
self--gravitating disc ($M_{\rm disc}=0.2 M$).  We explore the changes
of the binary eccentricity $e$, by simulating a sequence of binary
models that differ in the initial eccentricity $e_0$, only.  In
initially low-eccentric binaries, the eccentricity increases with
time, while in high-eccentric binaries $e$ declines, indicating the
existence of a limiting eccentricity $e_{\rm crit}$ that is found to
fall in the interval $[0.6,0.8]$.  We also present an analytical
interpretation for this saturation limit.  An important consequence of the
existence of $e_{\rm crit}$ is the detectability of a significant
residual eccentricity $e_{\rm LISA}$ by the proposed gravitational
wave detector {\it LISA}.  It is found that at the moment of entering
the {\it LISA} frequency domain
$e_{\rm LISA}\sim 10^{-3}-10^{-2}$; a signature of its earlier
coupling with the massive circumbinary disc.
We also observe large periodic inflows across the gap, occurring on
the binary and disc dynamical time scales rather than on the viscous
time. These periodic changes in the accretion rate (with amplitudes up
to $\sim 100\%$, depending on the binary eccentricity) can be
considered a fingerprint of eccentric sub-parsec binaries migrating
inside a circumbinary disc.

\end{abstract}


\section{INTRODUCTION}
\label{sec:Introduction}
Supermassive black hole (BH) binaries are currently postulated to form
in the aftermath of galaxy mergers \citep{begelman80}, despite the
difficulties, still present, in identifying them
observationally \citep[see][for a review]{colpid2009}.  Thanks to
advances in N-Body/hydrodynamical simulations, it has been shown that
major mergers of gas-rich disc galaxies with central black holes are
conducive to the formation of eccentric BH
binaries \citep[e.g.][]{Mayer07}.
Orbiting inside the massive gaseous nuclear disc resulting upon
collision, the two BHs continue to lose orbital energy and angular
momentum under the large-scale action of gas-dynamical friction, and
end up forming a circular Keplerian binary, on parsec
scales \citep[]{Escala2005,Dotti07,Dotti2009b}.
  As the gaseous and stellar
mass content inside the BH orbit continues to decrease in response to
the hardening of the binary, further inspiral is believed to be
controlled by the action of either three-body scattering of
individual stars and/or the interaction of the binary with a
circumbinary gaseous disc \citep[e.g.][]{MerrittReview05,Armitage:2002}.

The gravitational interaction of the massive BH binary with the
gaseous disc is believed to be of foremost importance to assess not
only its observability on sub-parsec scale, but its fate.
Gravitational waves start to dominate the BH inspiral (leading to
coalescence) only at tiny binary separations, of the order of a few
milli-parsec for a binary of $M \approx 10^6 \msun$.
If a viscous disc is present, Lindblad
resonances can cause BH migration down to the
gravitational wave (GW) inspiral domain
\citep[e.g.][]{GoldreichTremaine80,papaloizou77}. Following this
proposal, a number of studies have modelled BH migration in
Keplerian, geometrically thin $\alpha$-discs
\citep[]{Ivanov99, Gould2000, Armitage:2002,
Haiman2009, Lodato2009}.

Using high resolution hydrodynamical simulations, \citet{Jorge09}
recently investigated the evolution of the orbital elements of a
massive BH binary, under the hypotheses (i) that the binary, at the
radii of greatest interest (tenths of a parsec), is surrounded by a
self-gravitating, marginally stable disc, and (ii) that the binary has
excavated in its surroundings a cavity, i.e. a hollow density region
of a size nearly twice the binary orbital separation, due to the
prompt action of the binary's tidal torques.  The simulations
highlight one key aspect: that of the {\it increase} of the binary
{\it eccentricity}, $e$, during the decay of its semi-major axis.  The
excitation of $e$ was already noticed and studied
in \citet{Armitage:2005}, who investigated BH orbital decay in the
presence of a Keplerian $\alpha$-disc in two dimensions, as well as
in earlier analytical work by \citet{GoldreichSari03}
in the context of type-II planet migration.

The increase of $e$ has a number of interesting consequences.  First,
for a given semi-major axis, binaries with larger $e$ will lose energy
substantially faster via GWs, coalescing on a shorter time scale
\citep{Peters:1963ux}.
Second, accretion streams that leak through the cavity and fuel the
BHs happen with a better defined periodicity in the case of eccentric
binary \citep[e.g.,][]{arty96}
likely increasing the chance of BH binary
identification through AGN time-variable activity.
Finally, more eccentric binaries will retain
some residual eccentricity when detectable by the Laser Interferometer
Space Antenna ({\it LISA}) \citep{Berentzen2009,Amaro2010,Sesana2010}.
For these reasons it is important to
understand if, under disc-driven migration, the eccentricity keeps on
growing up to $e\approx 1$, or if there is a limiting eccentricity
toward which the binary orbit tends.

In this paper, we explore the binary--disc interaction with high
resolution N-body hydro simulations, modelling the circumbinary disc
as in \citet{Jorge09} (see Section~\ref{sec:Set_up}).  However,
instead of starting with binaries with low eccentricities, we now
construct a sequence of binaries with fixed semi--major axis, BH and disc--BH mass
ratios but with different initial eccentricities $e_0,$ varying it
from 0.2 to 0.8.  The binaries interact with a self-gravitating disc
changing their orbital elements.  With this approach
we assess whether the
eccentricity growth saturates, and at which value. We present a simple
analytical interpretation of our numerical results in
Section~\ref{sec:saturation}.  If the saturation eccentricity is
large, then the binary may reach coalescence with some residual
eccentricity, after GW emission has reduced it considerably.  This
issue was already discussed in
\citet{Armitage:2005} as a possible discriminant between
gas-driven versus stellar-driven inspiral.  In
Section~\ref{sec:parameter_estimation} we revisit this question in
detail, in the context of the proposed {\it LISA} mission.  The
simulations also provide information on gas streams that leak through
the cavity.  We investigate how the variability properties of the
accretion rate on to the BHs depend on the binary eccentricity.  This
analysis may lead to the identification of BH close binaries and
estimates of their orbital elements
(Section~\ref{sec:parameter_estimation}).

\section{Simulation Set-up}
\label{sec:Set_up}

\subsection{The Model}
\label{subsec:model}

We model a system composed of a binary black hole surrounded by a
gaseous disc.  Since we are interested in the sub-pc separation
regime, we assume that the binary torque has already excavated an
inner cavity in the gas distribution.  We also assume that the cooling
rate is long relative to the dynamical time scale, preventing disc
fragmentation \citep[e.g.][]{Rice05}.  We consider a binary with an
initial mass ratio\footnote{Unless otherwise stated, subscripts $1$
  and $2$ refer to the primary (more massive) and secondary (less
  massive) black hole, respectively.}  $q=M_2/M_1 = 1/3$ and a disc
with an initial mass $M_{\rm disc} = 0.2M$, where $M = M_1+M_2$ is the
total mass of the binary.  The binary has initial eccentricity $e_0$,
semi-major axis $a_0$, initial dynamical time $t_{\rm dyn}=f_0^{-1}=2
\pi / \Omega_0$, where $\Omega_0 = (GM/a_0^3)^{1/2}$.  Both the binary
and the disc rotate in the same plane and direction as expected from
the simulations of \cite{Dotti2009b}.  The disc is initially
axisymmetric, and extends from $2a_0$--$5a_0$.  Its initial surface
density profile is given by $\Sigma(R) \propto R^{-1}$, where $R$ is
the distance to the centre of mass of the system.

\subsection{Early Evolution}

\cite{Jorge09} modelled the evolution of low-eccentricity binaries in
the system discussed above.
They found that self-gravity drives the initially
uniformly distributed gas into a ring-like configuration located at $R
\approx 3 a_0$.  This ring eventually collapses and later spreads
again in roughly the same radial range it had in the initial
conditions ($2a_0$--$5a_0$).  However, instead of having a uniform
density distribution, the disc displays a clear spiral pattern.
\cite{Jorge09} found that this configuration remains stable for at
least $3000 \Omega_0^{-1}$, and that during this time the binary both
shrinks and gains eccentricity due to its interaction with the disc.
In this study, we skip the early transient evolution and start from a
snapshot
taken at $t=500 \Omega_0^{-1}$.  At this time,
the disc has already settled into the steady-state configuration.

\subsection{The New Simulations}
\label{sec:new}

Our goal is to study the secular evolution of the binary--disc system,
focusing in the evolution of the binary eccentricity.  The ideal
method would be to follow the binary from an initial, pc-scale
separation, until it reaches the GW-dominated regime.  Unfortunately
such an approach is not feasible.  The time scale for decay is $\sim 10^4
\Omega_0^{-1}$ \citep{Jorge09}, much longer than what we can feasibly simulate
with current computational power.  Moreover, as the binary
shrinks, its angular momentum is transferred to the disc.  Without
appropriate boundary conditions, this results in the unphysical
expansion of the disc, slowing further the evolution of the system
\citep{Jorge09}.
To accomplish our goal we take an indirect approach. We run a set of
simulations where the gas configuration was taken from the steady state of a previous simulation,
as described above, but the
binary had different initial eccentricities. The energy of the binary was conserved, i.e.
its semi--major axis $a$ was fixed, only the angular
momentum of the binary was changed to accomplish the various initial eccentricities $e_0$.
We then extrapolate the long-term evolution of the eccentricity
interpreting the results of the different runs as snapshots of the
binary life taken at different ages.

\subsection{Numerical method}
\label{subsec:numerical_method}

\begin{figure}
  \hspace{-0.9cm}
  \includegraphics[width=1.05\linewidth]{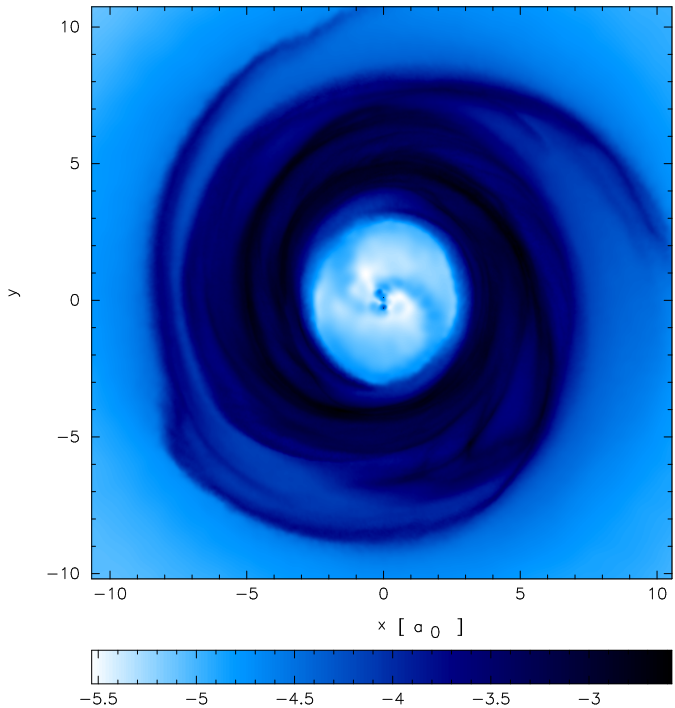}
  \caption{Face-on view of the circumbinary disc surrounding a BH binary of initial eccentricity $e_0=0.6$
  after 180 orbits. The gas density is colour-coded on a logarithmic scale with brighter colours
  corresponding to lower gas density; axes in units of $a_0$.  The figure shows the
 spiral patterns excited in the disc, the gap surrounding the binary,  and
 the yin-yang shaped gas inflows around the BHs. Figure made using {\sc SPLASH} \citep{Price2007} }
 \label{fig:sim}
\end{figure}

To simulate the binary--disc system, we use the numerical method
described in detail by \citet{Jorge09}.  We use a modified version of
the SPH code {\sc Gadget-2} \citep{Springel05}.
We allow the gas to cool on a time scale which is proportional to the
local dynamical time of the disc.  To prevent it from fragmenting, we
set $\beta = t_{\rm cool}/t_{\rm dyn} = 10$.  Unlike \cite{Jorge09},
we assume that the small amount of gas present in the inner cavity ($r
\lsim 1.75a$) is isothermal, with an internal energy per unit mass $u
\approx 0.14 (GM/R)$.  The effect of this recipe is to confine the gas in
the inner region to a relatively thin geometry.
The gravitational interaction between particles is calculated with a
Barnes--Hut tree. For all runs we use 2 million particles, a number which
has been shown to be sufficient by \citet{Jorge09}. Since we are interested in following the
evolution of the binary orbit accurately, we take the BHs out of the tree
and compute the gravitational forces acting on them directly, i.e.
summing up the contributions from each gas particle.  Moreover, to
ensure an accurate integration, the dynamics of the BHs is followed with a fixed
time-step, equal to 0.01 $\Omega_0^{-1}$.
The BH binary is modelled as a pair of point masses, and their
potentials are assumed to be Newtonian.  Relativistic corrections,
important only when the binary separation decays below $\sim 2$
mpc \citep{Peters:1963ux}, are not included in the SPH simulations but
are considered in Section~\ref{sec:parameter_estimation}, when estimating
the eccentricity of binaries entering the LISA band.  Gas particles
approaching either BH are taken away from the simulation in order to
avoid the very small time-steps they would require.  They are
considered to be accreted, and  their mass and momentum are transferred
to the corresponding BH \citep{Bate95, Cuadra2006a}.
In the present simulations the sink radius around
each BH, below which particles are accreted, is
set to $0.03 a_0$. A face on  view of the disc surface density is
shown in Fig.~\ref{fig:sim} in which the gas has already
relaxed around a binary of $e_0=0.6$. It shows the typical spiral
arms in the disc and the resonant streams in the inner gap region.


\begin{figure}
  \hspace{-0.9cm}
  \includegraphics[width=1.15\linewidth]{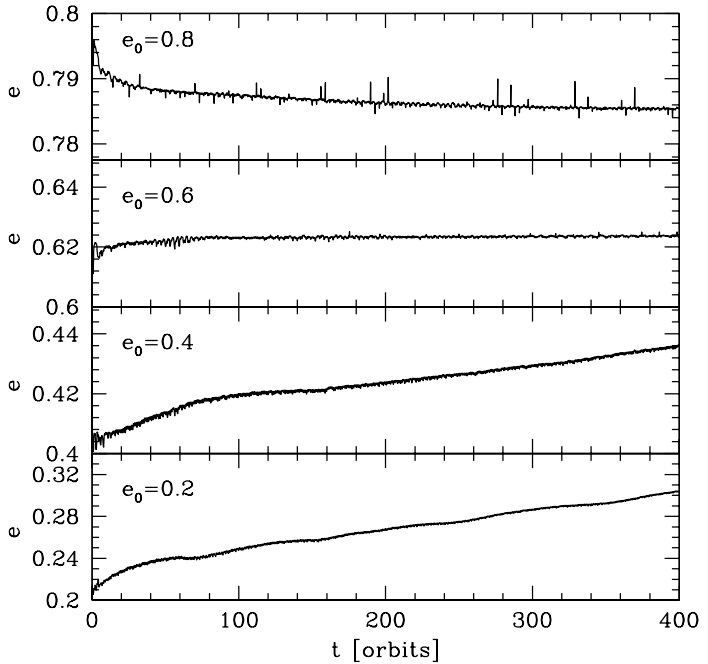}
  \caption{The eccentricity evolution of the four standard runs, starting from $e_0=0.2,0.4,0.6,0.8$ bottom to top. }
 \label{fig:eccall}
\end{figure}

\section{Eccentricity evolution}
\label{sec:simulations}
 As described in Section~\ref{sec:new}, we prepared four initial
 conditions identical but for the initial values of the binary
 eccentricity.  In Fig.~\ref{fig:eccall} we show the evolution of $e$
 for four runs with initial eccentricities $e_0 =0.2,0.4,0.6,0.8$,
 respectively (bottom to top).  The bottom panel depicts the monotonic
 rise of $e$, for the run with $e_0=0.2$: The eccentricity increases
 almost linearly after the first $70\, f_0^{-1}.$ The run for
 $e_0=0.4$ (second panel) displays a similar behaviour, but the slope
 $de/dt$ is much shallower (note the different scales in the $y$ axes of
 Fig.~\ref{fig:eccall}). In the third panel, corresponding to
 $e_0=0.6$, we observe a fast increase of the eccentricity up to
 $e=0.62$ within the first few orbits; afterwards the eccentricity
 saturates, approaching a constant with $de/dt \sim 0^+$.  The top
 panel refers to the run with the largest initial eccentricity
 explored, $e_0=0.8$. This time, the eccentricity exhibits a negative
 slope with $d^2e/dt^2$ steadily decreasing until $de/dt \sim 0^-$.

The key result, illustrated in Fig.~\ref{fig:eccall}, is the existence
of a limiting $e_{\rm crit}$ that the BH binary approaches in its
interaction with the disc.  Since the runs were halted after 400
orbital cycles, we can only bracket the interval in which $e_{\rm
  crit}$ lies: $e_{\rm crit} \in [0.62,0.78]$.  The reason of this
uncertainty is technical as we find that the decline of $e$ is very
hard to follow numerically due to the fast expulsion of the gas out of
the region where torques can still effectively interact with the BH
binary -- an effect that increases with the binary eccentricity, as expected.
Indeed, if we define $R_{\rm gap}$ as the inner location of the disc's
half-maximal surface density, we find that the gas moves from an initial value
of $R_{\rm gap} \approx 2 a_0$ to a time-averaged value of $\approx
2.6 a_0$, $3.0 a_0$, $3.4 a_0$, and $3.8 a_0$ during the first 53 binary
orbits, for the runs with an initial binary eccentricity of 0.2, 0.4, 0.6, and 0.8,
respectively. Such an expansion of the gas is not
unexpected since no outer inflow boundary
conditions were implemented in our simulations.
While the rate of eccentricity change is affected by the expansion
resulting from the initial orbital set-up, its long-term trend
(whether it increases or decreases) is a robust conclusion from our
numerical study.\\

Our simulations strongly suggest the existence of a saturation in the
disc-driven eccentricity growth, but do not
pinpoint the exact value of $e_{\rm crit}$. In the next section we
discuss the physical reasons for this limit and analytically predict
the value of $e_{\rm crit}$.


\section{Explanation of the saturation}
\label{sec:saturation}

\begin{figure*}
  \centering
   \includegraphics[width=0.48\textwidth]{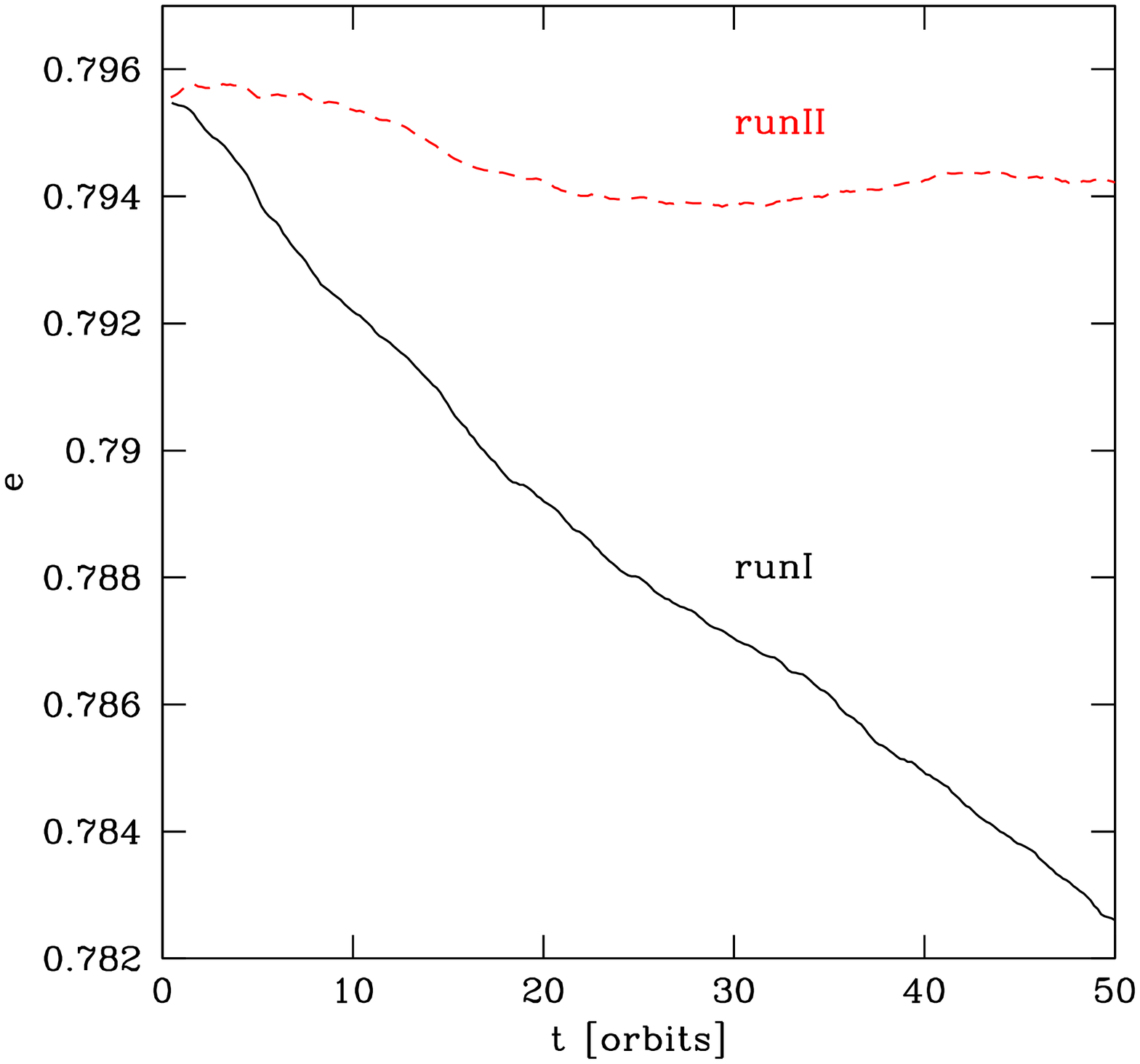}
   \includegraphics[width=0.48\textwidth]{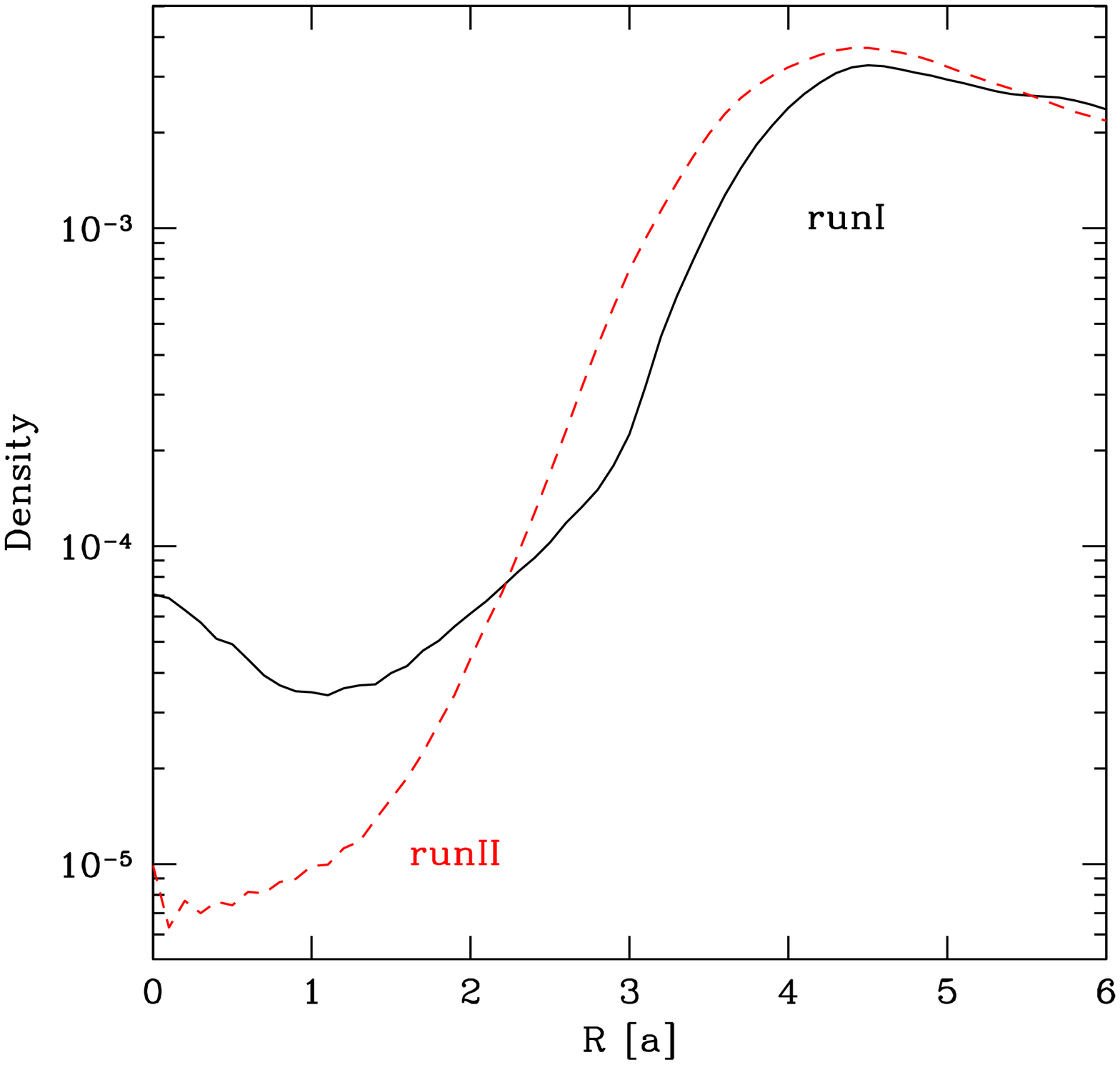}
\caption{Additional high eccentricity runs where the  initial semi-major axis is
reduced by a factor 1.8, compared to the default runs. These short runs are used to test
the emerging picture of a limiting eccentricity depending on the amount of streams present in the cavity. The difference between
the two runs is the thermodynamical treatment of the gas inside the cavity. For runI this is identical to the default runs whereas
in runII we suppress gaseous inflows into the gap.
 Left panel: eccentricity versus time, for $e_0=0.8$.
Solid line refers to runI, while dashed line to runII.
Right panel: azimuthally averaged disc surface density as a function of $R[a]$ in arbitrary units.
Surface density is averaged over the orbits $20-30$.
Solid line refers to runI, dashed line to runII (see text for description).}
\label{torques}
\end{figure*}

The {\it growth} of the eccentricity, from initial values $e_0$ below
a critical eccentricity $e_{\rm crit}$, and the {\it decline} of $e$
from initial values $e_0>e_{\rm crit}$ call for a simple physical
interpretation.  The increase of the eccentricity caused by the
interaction of the binary with an external disc is a known fact for
very unequal binaries where the non-axisymmetric potential
perturbations are small \citep[as in the case of planetary migration,
see e.g.][]{GoldreichTremaine80, GoldreichSari03,
Armitage:2005}.

\cite{GoldreichTremaine80} have shown that in the high mass-ratio limit the binary-disc
transfer of angular momentum occurs secularly through torques excited
in the disc by the binary at discrete Lindblad and co-rotation
resonances.  Damping and/or growth of $e$ thus depends on the relative
importance of these opposing torques (and so on how fluid elements are
distributed in the disc).  Principal Lindblad resonances are known to
be responsible for opening a gap in the disc. As a consequence of disc
clearance, co-rotation and inner Lindblad resonances are reduced in
power.  This consideration led \cite{GoldreichSari03} to show that
only the outer Lindblad resonances, remaining after gap opening, cause
the increase of the eccentricity for initially low eccentric
binaries.

For the comparable mass limit studied in this paper ($q=1/3$)
we have a simpler explanation.
 An initially small $e$ increases because of the larger deceleration
experienced by the secondary BH near apo-apsis with respect to
peri-apsis \citep[see e.g.,][]{Lin79,Artymowicz91}. The longer time spent when nearing apo-apsis, and the
larger over-density excited in the disc by the hole's gravitational
pull due to its immediate proximity are both conducive to a net
deceleration of the hole that causes the increase of the binary
eccentricity.  This increase continues as long as the secondary BH has
a larger angular velocity at its apo-apsis $\omega_{\rm 2, apo}$ than
the fluid elements in the disc $\omega_{\rm disc}$.  When this
reverses, the density wake excited by the BH moves ahead imparting to
the hole, near apo-apsis, a net tangential acceleration that tends to
increase the angular momentum content of the binary, decreasing $e$.
This argument is valid if the disc and the binary angular momenta
are aligned. If they are antialigned (i.e. for a retrograde disc) the
interaction between the BHs and the gas increases the eccentricity up to
$e \approx 1$ \citep{Nixon2011}, resulting in a fast coalescence of
the binary.  We limit our investigation to discs corotating with the
binary, as expected if they form together during a gas rich galaxy
merger \citep{Mayer07,Dotti2009b}. In this case the torques on the
secondary will be minimal if $\omega_{\rm disc} = \omega_{\rm 2,
apo}$.
Approximating the binary as a purely Keplerian system and the gaseous disc to be in
Keplerian motion around a mass $M_1+M_2$ located at the system center of mass (COM),
it is easy to derive:  
\begin{eqnarray}
\omega_{\rm 2, apo}^2&=& \frac{G M_1 (1+q)}{(1+e)^2 a^3} \left( \frac{2}{(1+e)}-1\right)\\
\omega_{\rm disc}^2&=& \frac{G (M_1+M_2)}{R_T^3},
\end{eqnarray}
where we defined $R_T$ to be the distance of the
strongest torque on the binary as  measured from COM.
Equating $\omega_{\rm 2, apo}^2=\omega_{\rm disc}^2$ yields:
\begin{equation}
\frac{1}{R_T^3} =\frac{1 }{(1+e)^2 a^3} \left( \frac{2}{(1+e)}-1\right),
\end{equation}
which can be rearranged as
\begin{equation}
\delta^3 = \frac{(1+e)^3 }{(1-e)},
\label{eq:limit1}
\end{equation}
with $\delta=R_T/a$.
Eq.~(\ref{eq:limit1}) implies the existence of a limiting eccentricity
$e_{\rm crit}$ that we can infer via numerical
inversion of Eq.~(\ref{eq:limit1}). The expression
\begin{equation}
e_{\rm crit}=0.66\sqrt{{\rm ln}(\delta-0.65)}+0.19
\label{eq:ecrit}
\end{equation}
provides an analytical fit to the result within a $2\%$ accuracy in the
range $1.8<\delta<4.5$, relevant to our study assuming that, in a first approximation,
 $\delta$ can be set equal to the inner edge of the disc, $R_{\rm gap}$.
Note that in this derivation, {\it for a fixed } $\delta$, $e_{\rm crit}$ is
independent of the binary mass ratio. To compare the predictions of this toy model
with the simulations we need to define the inner edge of the disc. This is somewhat tricky
since the disc profile is not a step function at a certain $R/a$. In our initial simulation
the clean region within the gap has a size of $R_{\rm gap}\approx 2 a$. At larger distances the disc
density increases reaching a maximum around $R/a\approx 2.5$. For $2<\delta<2.5$ we
get $0.55<e_{\rm crit}<0.69$
which is within the range obtained from the numerical simulations described in Section~\ref{sec:simulations}.
 Note that Eq.~\ref{eq:ecrit} depends on the specific value of $\delta$, i.e. on how close inflows of gas can get to the binary.
Even though $\delta$ can in principle be measured from our simulations, its value would also be affected by the lack of physical outer-boundary conditions.
Instead, $\delta$ is usually determined equating the viscous torque in the accretion disc with the positive torque exerted by the binary \citep[see, e.g. Eq.~15 in][]{Artymowicz1994}.
\cite{Artymowicz1994} found that the size of the gap depends on $e$.
For $e\approx 0.6$, $q=0.3$, disc aspect ratio $H/R = 0.03$ and a viscous parameter $\alpha =0.1$,
they predict $R_{\rm gap} \approx 2.9 a$, corresponding to a 5:1 commensurability resonance.
Using this value for $\delta$ we would obtain a larger value of $e_{\rm crit}\approx 0.77$.
Note that the interaction between the binary and the disc becomes less efficient as the disc expands whereas the gravitational pull of the
 tenuous gas onto the secondary at peri-apsis increases. So even in a system where the influence of the gas inside the cavity is completely negligible,
 it is not clear if the binary could reach such a high $e_{\rm crit}$ on a relevant time scale.
Note that a retrograde disc would not expand, since the
interaction with the binary decreases its angular momentum. In this
case the eccentricity growth remains efficient up to $e \approx 1$
\citep{Nixon2011}.

A direct comparison between our results and
\cite{Artymowicz1994}'s prediction is not
straightforward. Although our self gravitating disc is
able to redistribute angular momentum efficiently, its total amount has to
be conserved. Thus, discs hosting very eccentric binaries ($e=0.6, 0.8$)
keep on expanding after a short impulsive interaction with the binary (as
discussed in Section~\ref{sec:simulations}). The interaction between the
disc and the binary is extremely inefficient when $R_{\rm gap}\gtrsim 4a$ (see
the two top panels in Fig.~2). Therefore, although a larger $R_{\rm gap}$, in
first approximation, implies a larger $\delta$ implying a larger $e_{\rm
crit}$, it also results in longer timescales for the eccentricity
evolution.

The feeding of a BH binary forming in a gas rich galaxy merger can
be a very dynamic process, and the interaction with a single
circumbinary disc could be too idealized a picture. Larger scale
simulations show episodic gas inflows due to the dynamical evolution
of the nucleus of the remnant \citep[see
e.g.][]{Escala2006,Hopkins2010}.  In this scenario the binary can
still interact with a disc and excavate a gap, but the size of it
would be time dependent (as in the simulations presented here) and
would also depend on the angular momentum distribution of the
inflowing streams, resulting in a range of $e_{\rm crit}.$

\subsection{	Testing the emerging picture}
Eq.~\ref{eq:ecrit} shows that $e_{\rm crit}$ depends on the location of the strongest torque
$\delta$ and thus, in first approximation, on the size of the gap $R_{\rm gap}$.  In order to cross-check our results, we
performed two additional simulations of the $e_0=0.8$ case, in which
$a(1+e)$ was kept fixed, reducing the semi-major axis by a factor
$1/1.8$, thus increasing the {\it relative} gap size $R_{\rm gap}$ by
$80\%$.  These runs simulate a situation where the infalling material
stays at a large distance from the eccentric binary and does not reach
$R_{gap}\approx 2.5 a$, typical for the low eccentricity cases presented
above.

The analysis performed in the previous section only accounts for
the pull of the disc when the secondary is at apo-apsis,
neglecting torques exerted by the infalling material forming mini-accretion
discs around the two BHs . For low $e_0,$ this approximation works well,
because the separation of the two BHs is always much larger than the size
of the inner mini-discs. However, in the high $e_0$, small $a$ case tested here,
the secondary BH, at each peri-apsis passage, experiences a significant
drag onto the inflowing mass accumulating around the primary. Such drag causes the circularization of the orbit.
Therefore, the secular evolution of the binary is determined by two
factors: i) the distance of the gap from the secondary BH at apo-apsis, and ii)
the amount of inflowing gas through the gap onto the primary BH.

In order to separate the two effects we set two simulations with
identical initial conditions as described above. In runI,
we keep exactly the thermodynamics employed in our fiducial runs, that
allow a stable accretion mini-disc to form around the primary hole;
in runII the gas inside the gap evolves with
the $\beta$--cooling enabled just as in the rest of the disc, and can be
heated by adiabatic compression.
This suppresses the gaseous inflows into the cavity and prevents the gas from forming a significant circumprimary disc.
As shown in the left panel of Fig.~\ref{torques}, runI experiences a
substantial steady decline in $e$, whereas in runII, after a slight initial
reduction, the eccentricity stays more or less constant.
Such a result confirms our understanding of the dynamics of the system.
In runI the secondary encounters the high density region formed around the
primary at each peri-apsis passage and is slightly decelerated onto
a more circular orbit. In runII, after a short initial relaxation phase,
there is not enough gas in the center to cause further circularization
(compare the two central densities in the right panel of Fig.~\ref{torques});
on the other hand, the gap is large enough for the disc--binary interaction
to be weak and, therefore, the eccentricity growth to be very inefficient.
Note also, that for a wider binary the same effect holds, however, only
if the secondary passes through the mini--disc of the primary, the size of which is independent of $a$.
 That's why in comparison to the default runs in Section~3,
the effect is visible more clearly here in the case of the narrower binary.
Thus, the predicted limiting eccentricity $e_{\rm crit} \approx 0.88$ expected
for $\delta=3.5$ (approximately the size of the gap in these close--separation
simuations) can not be achieved.


Although the torques exerted by the inside-cavity material when the binary
eccentricity is high add complexity to the emerging picture, this strengthen
the result of a limiting eccentricity in the range $0.6<e< 0.8$ for the
BH binary-disc configurations examined in this paper.




\section{Observational consequences}

\label{sec:parameter_estimation}
In this Section we focus on the impacts that our findings might have
on the long-standing search for close BH binary systems in the
Universe.  First, we investigate possible periodicities residing in
the accretion flows onto the two BHs enhancing our ability to identify
such elusive sources. Then, we study the influence of a high limiting
eccentricity (attained during migration) on future gravitational wave
observations with {\it LISA}.

\subsection{Periodically modulated accretion flows}
\label{sec:accretion}

\begin{figure}
 \includegraphics[width=1\linewidth]{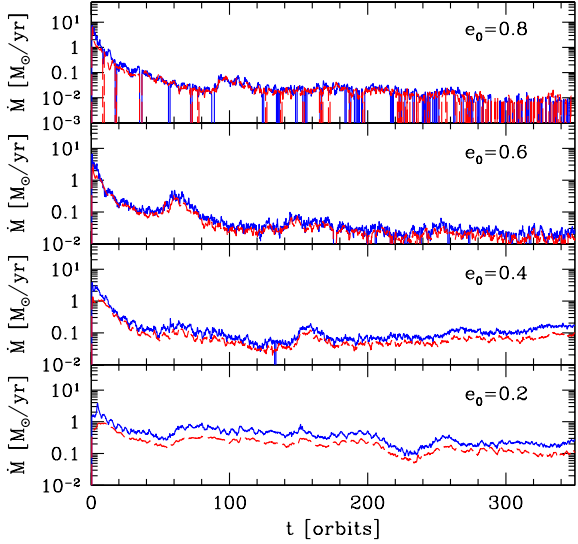}
 \caption{Mass accretion rates onto the BHs, for the runs with $e_{0}= 0.2,0.4,0.6,0.8$ (bottom to top).
 Dashed (\textcolor{red}{red}) line refers to the primary BH, solid (\textcolor{blue}{blue}) line to the
 lighter secondary hole.}
 \label{fig:08eccruns}
\end{figure}

Fig.~\ref{fig:08eccruns} shows the evolution of the accretion rate
\textcolor{red}{${\dot M}_1$} and \textcolor{blue}{${\dot M}_2$} onto
each hole, for the four runs with $e_0=0.2,0.4,0.6,0.8$. In order to
interpret our results we consider the binary to have a total mass
$M=3.5 \times10^6 M_{\odot}$, typical for expected {\it LISA} detections,
 and an initial semi-major axis $a_0=0.038$
pc. Under this assumption, for a radiative efficiency of 0.1, the
Eddington limit would correspond to accretion rates
\textcolor{red}{${\dot M}_{1,\rm E} =0.06 M_{\odot}\,\rm {yr}^{-1}$}
and \textcolor{blue}{${\dot M}_{2, \rm E} =0.02 M_{\odot}\rm
  {yr}^{-1}$}. Fig.~\ref{fig:08eccruns} shows that this limit is
fulfilled for the two high eccentricity runs only, whereas for the
low-$e$ runs the BHs accrete at super Eddington rates. This is
possible since the numerics do not include any radiative feedback.
The accretion rates drop significantly in the runs with initially
higher eccentricity, owing to the expansion of the gap size with time
(as discussed in Section 3).

\begin{figure*}
  \centering
  \includegraphics[width=0.48\textwidth]{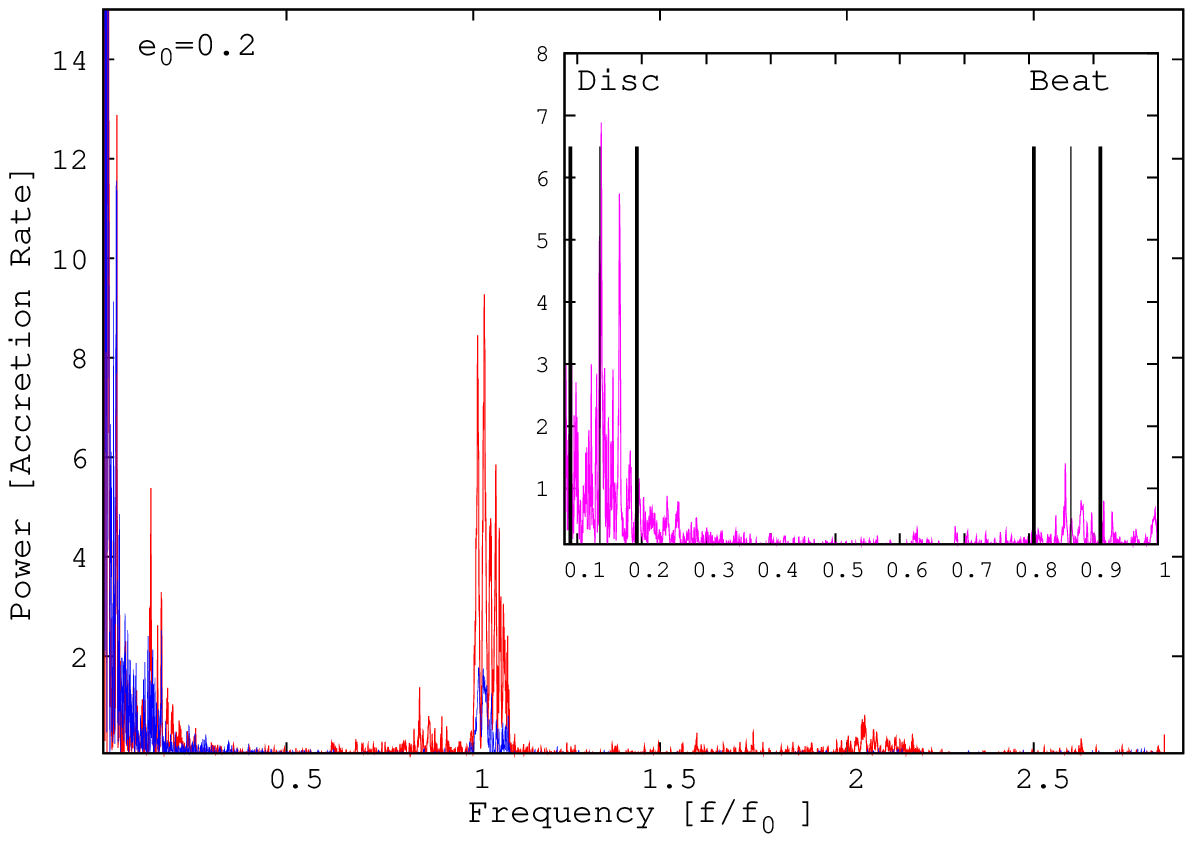}
  \includegraphics[width=0.48\textwidth]{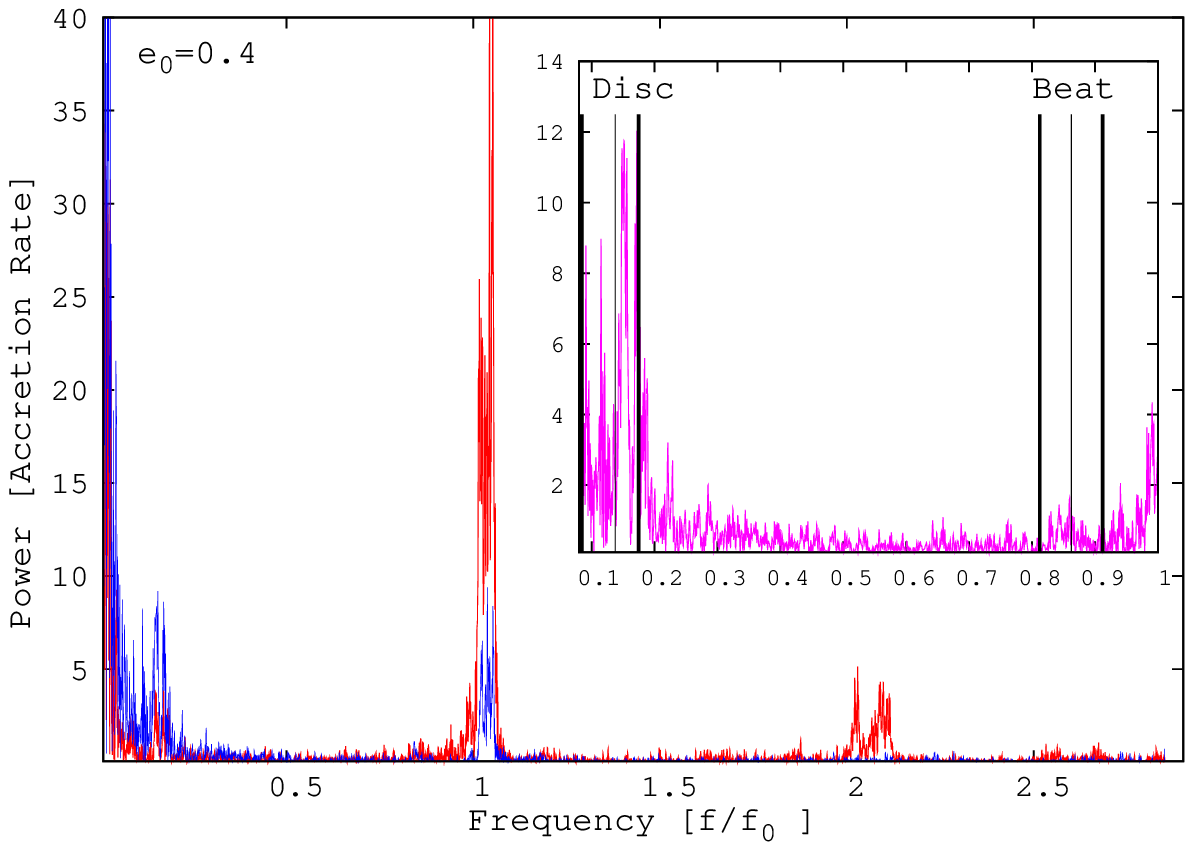}
  \includegraphics[width=0.48\textwidth]{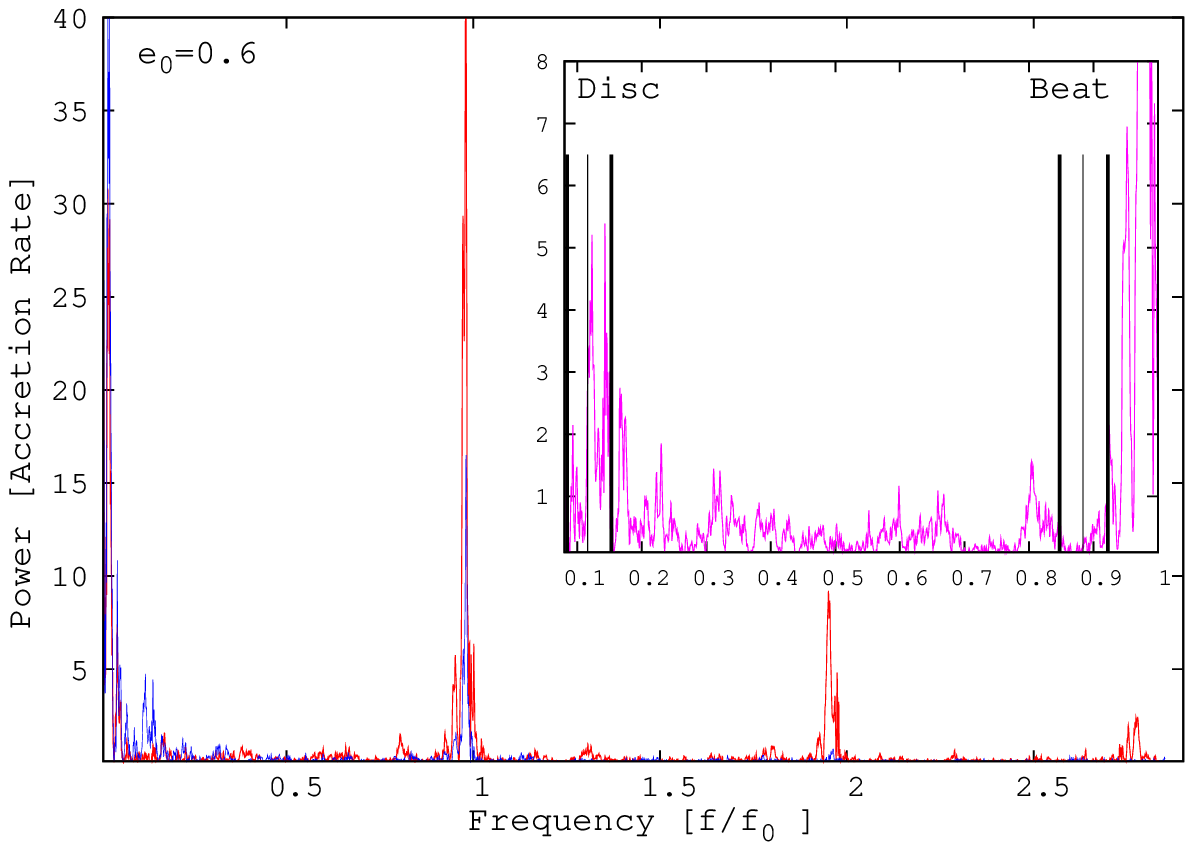}
  \includegraphics[width=0.48\textwidth]{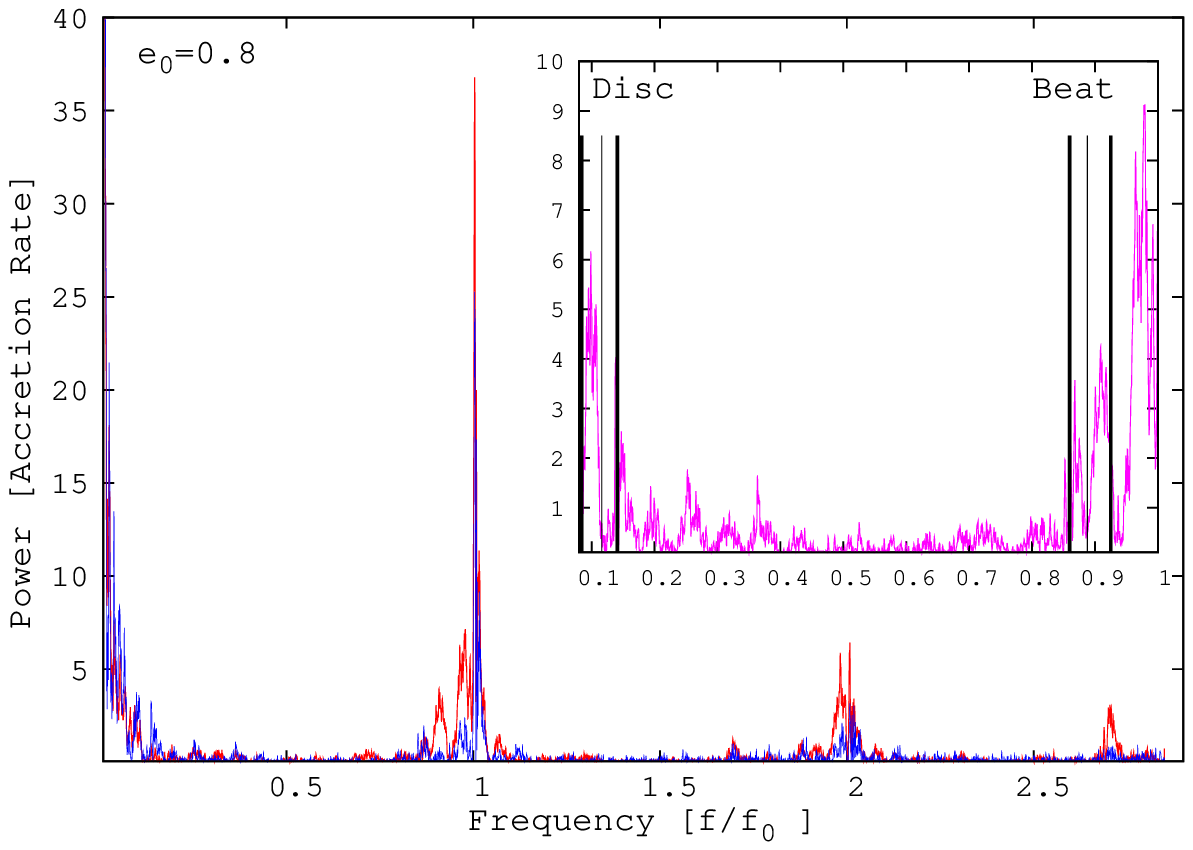}
  \caption{Power spectrum of the accretion rate (in arbitrary units$^2$)
    onto the primary (\textcolor{blue}{blue}) and secondary
    (\textcolor{red}{red}) BHs.  Frequencies are in units of the
    initial binary orbital frequency $f_0$. The inlays show zoom-ins of the
    power spectra, in the frequency range $0.1$--$1 f_0$ computed summing
    the accretion rate from the two BHs (\textcolor{RubineRed}{pink}).
    The expected intervals for the disc and the beat frequencies are
    marked by the thick vertical black lines, as labelled in the
    Figure.  }\label{fig:acchigh}
\end{figure*}

Fig.~\ref{fig:acchigh} shows the power spectra of the accretion rates
${\dot M}$ onto the two BHs.  The frequency $f$ (on the $x$-axis) is
in units of the binary orbital frequency $f_0$, and the power spectral density in
arbitrary units.  A clear periodicity emerges at the orbital frequency
$f_0$, indicating a modulation of the inflow rate, induced by the
orbital motion \citep{arty96,haya08}. Note that smearing of the peaks
at $\sim f_0$, in Fig.~\ref{fig:acchigh}, for $e_0=0.2$ and $0.4$, is
due to the few-percent shrinking of the semi-major axis,
and therefore also of the orbital period, during the
evolution.  As the binary eccentricity increases, the second and third
harmonics of the orbital frequency increase in power and become
visible.  In the inlay of each panel the power spectrum associated
to the total mass transfer rate onto the binary is plotted in the frequency
range $0.1 f_0< f< 1.0 f_0$ to illustrate the presence of other
characteristic features at: (i) the frequency associated to the
rotation of the fluid in the dense part of the disc: $f_{\rm
  disc}/f_0=(a_0/r_{\rm disc})^{3/2}$; (ii) the beat frequency,
i.e. the difference between the binary and the disc rotation
frequencies: $f_{\rm
  beat}/f_0=1-(a_0/r_{\rm disc})^{3/2}$.  Here $r_{\rm disc}$ denotes
the radial distance where the disc surface density has its maximum.
Since the disc has a broad density profile, we consider the two values
$r_{-}$ and $r_{+}$ defined by the full width half maximum (FWHM) of
the density and use those to estimate the expected disc and beat
frequency intervals (enclosed by the two pairs of thick black lines in
the inset of Fig.~\ref{fig:acchigh}). As expected, we observe broad
features consistent with the predicted frequency ranges. Signatures of the disc are
always visible in these plots, with a complex line structure mirroring the
over-densities in its spiral arms. The beat is very distinct in the
$e_0=0.8$ run, marginally visible the other runs.
We further notice that the significance of the peak \footnote{We utilize the normalized Lomb--Scargle periodogram here, wherein the significance of each peak
is directly given by the false-alarm probability (FAP) \citep{scargle82}. Since the number of independent frequencies is the same for all four runs, the FAP scales identically for all runs, thus the relative height translates into significance. For our runs, a peak needs to exceed a height of $12$ in order to have a FAP of $0.01$}
 in the power
spectrum, at the binary orbital frequency $f_0$, is weaker for
low-eccentric binaries ($e_0=0.2$) than for binaries with higher
eccentricities. This agrees with previous works \citep[cf.][]{Jorge09}
that show a mild periodicity in the accretion rate in the case of
quasi circular binaries.  Thus, a periodic signal is expected to be a
distinctive signature of eccentric massive BH binaries.

The presence of periodicities in the accretion flows opens interesting
prospects for monitoring sub-parsec BH eccentric binaries in
circumbinary discs. Our fiducial system has $M=3.5 \times10^6\,{\rm
  M}_{\odot}$ and an initial semi-major axis $a_0=0.038\,$pc
corresponding to an orbital period of $348$ years, exceeding a human
lifetime. Since the binary fingerprints in the accretion rates are
related to the dynamical time, we can extrapolate our results to
smaller periods as long as the disc and the binary are dynamically
coupled (see next Section). For a binary with $M=3.5\times10^6 \, {\rm
  M}_{\odot}$ and $q=1/3$, binary-disc coupling may survive down to
much shorter periods of $\sim 1$ month, making the observation of such
periodicities astrophysically feasible.  The interval of
modulation $\Delta ({\dot M})$ from our runs is at the level of:
$\Delta ({\dot M}) \in [10,50] \%$ for $e_0=0.2$, $\in [40,100] \%$ for
0.4, $\in [40,90] \%$ for 0.6, and  $\in [10,50] \%$ for 0.8.
Assuming a luminosity
proportional to the time dependent accretion rate, a
periodic monitoring of such sources will allow to
construct the light curve for several years. An amplitude modulation
of up to $100\%$ over 10-to-100 cycles will thus be easily
identifiable.

\subsection{Residual eccentricity in GW-observations}
\label{sec:GW}

The existence of a limiting eccentricity that is maintained during the
coupled evolution of the disc-binary system has important consequences
for the detection of the binary as GW source in the latest stage of
its evolution, i.e.  during the last year of GW inspiral towards
coalescence. Since the systems in our simulations are far from
coalescence (in our fiducial rescaling $a=0.038$pc, corresponding to
$\sim 10^5$ Schwarzschild radii of the primary hole), in the following
we will extrapolate our findings to much smaller scales (order of
$\sim 10^3$ Schwarzschild radii) making use of the standard optically
thick, geometrically thin $\alpha$-disc recipe \citep{ss73}.

In the standard picture of BH migration, the BHs reach closer
separations under the action of viscous torques exerted by the
circumbinary disc. This holds true as long as the migration time scale
$t_{\rm m}$ is shorter than the binary GW decay time scale $t_{\rm
  GW}$. Since the former scales as
$\propto a^{7/8}$ or $a^{35/16}$ for
gas and radiation pressure supported discs \citep[]{Haiman2009}, while
the latter as $\propto a^4$, there will eventually be a critical
separation $a_{\rm dec}$ below which GW emission takes over and the
binary decouples from the disc. After decoupling, binary-disc mutual
torques are ineffective and the binary evolution is driven by GWs
only. GWs tend to circularize the binary, but if decoupling occurs at
small $a$, there might not be enough room for complete orbit
circularization before entering the {\it LISA} frequency domain.  Even a
residual eccentricity as small as $e\sim10^{-4}$ may be easily
detectable \citep{jc10}, and it has to be accounted for, for a
trustworthy parameter estimation of the GW source
\citep{portersesana10}.

To estimate the residual eccentricity in the
{\it LISA} band $e_{\rm LISA}$ we need four ingredients:
\begin{enumerate}
\item the binary eccentricity at decoupling, $e_{\rm dec}$;
\item the binary semi-major axis at decoupling, $a_{\rm dec}$;
\item a model for the GW decay after decoupling;
\item an estimation of $f_{\rm LISA}$ at which $e_{\rm LISA}$ has to be
computed.
\end{enumerate}
Being interested in {\it LISA} BH binaries, we consider
systems characterized by $10^5\msun<M_1<10^7\msun$ and $0.01<q<1$.
Item (i) is directly extracted from the simulations and the analytical
argument presented in this paper. We assume that, at decoupling, the
binary has the limiting eccentricity $e_{\rm crit}$ given by equation (2).

Because of its small extent, the circumbinary disc assumed in our
simulations is unable to transfer the binary angular momentum outwards
efficiently for a prolonged time scale. It is therefore unsuitable for
estimating a disc-driven binary decay rate to be compared to the GW
angular momentum loss. A viable short cut to compute $a_{\rm dec}$
(item (ii)) is to link our disc to a standard thin accretion disc and
to estimate the gas-driven migration time scale in that approximation.
When scaled to physical units, our binary has $a_0=0.038\,$pc. At such
a separation, the circumbinary disc can be described as a
steady-state, geometrically thin, optically thick Shakura-Sunyaev
$\alpha$ disc \citep[]{Haiman2009}.  Accordingly, the disc has a mass
\begin{equation}
M_{\rm d}=1.26\times10^3{\msun}\,\alpha_{0.3}^{-4/5}\left(\frac{\dot{m}}{\epsilon_{0.1}}\right)^{7/10} M_7^{11/5}(R_{\rm out}^{5/4}-R_{\rm in}^{5/4}),
\end{equation}
where $\alpha_{0.3}$ is viscosity parameter normalized to 0.3,
$\dot{m}=\dot{M}/\dot{M}_{\rm E}$ is the accretion rate (in units of
the Eddington rate), $\epsilon_{0.1}$ is the radiative efficiency
normalized to 0.1, $M_7$ is the total mass of the binary in units of
$10^7\msun$; the two limiting radii of the disc, $R_{\rm in}$ and
$R_{\rm out}$, are expressed in units of $ 10^3R_{\rm Sch}$ (with
$R_{\rm Sch}=2GM/c^2$) and correspond to $R_{\rm in}=2a_0$ and $R_{\rm
  out}=10a_0$, respectively.  With this choice we infer a total disc
mass $M_{\rm d}\sim0.25 M$ which is comparable to our relaxed disc.
In such a disc the time scale for migration of the secondary BH onto
the primary is given by (Eq. 26a of \citet{Haiman2009})

\begin{equation}
t_{\rm m} = 1.5\times10^5 \,{\rm yr}\, M_7^{5/8} q_{\rm s}^{3/8} {\tilde a}_3^{35/16},
\end{equation}
where now ${\tilde a}_3$ is the binary semi-major axis in units of $10^3R_{\rm Sch}$ and
$q_{\rm s}=4q/(1+q^2)$ is the symmetric binary mass ratio.
This time scale has to be compared with the GW decay time scale for an eccentric binary
which, in the quadrupole approximation, is given by \citep{Peters:1963ux}

\begin{equation}
t_{\rm GW}=a\frac{dt}{da}=7.84\times10^6 \,{\rm yr}\, M_7 q_{\rm s}^{-1} {\tilde a}_3^{4}F(e)^{-1},
\end{equation}
where
\begin{equation}
F(e)=(1-e^2)^{-7/2}\left(1+\frac{73}{24}e^2 +\frac{37}{96}e^4 \right).
\end{equation}
The disc-binary decoupling occurs when $t_{\rm GW}=t_{\rm m}$,
and this happens somewhere in the range of binary separations between
$a_{\rm dec}\sim 10^2-10^3 R_{\rm sch}$, depending on the binary mass
and mass ratio. From that point on the dynamics of the binary is
driven by GW emission, only.

To address point (iii), we integrate the Post Newtonian equation for
eccentric binaries given by \cite{junker92}, following the
eccentricity evolution down to the last stable orbit. {\it LISA} will be
sensitive to GWs in the frequency range $10^{-4}-0.1$ Hz and, in
general, it will be able to monitor the final year of the binary
evolution with high signal-to-noise ratio.  We therefore set (item
(iv)) $f_{\rm LISA}={\rm max}[10^{-4}{\rm Hz},f(1{\rm yr})]$, where
$f(1{\rm yr})$ is the GW frequency observed one year before the final
coalescence.  Note that the observed GW frequency is related to the
rest-frame emitted frequency $f_{\rm r}$ as $f=f_{\rm r}/(1+z)$. This
means that $e_{\rm LISA}$, defined as the eccentricity of the BH
binary at the time of entrance in the {\it LISA} band, depends on the source
redshift. The $10^{-4}{\rm Hz}$ cut-off in observed frequency
corresponds to higher emitted frequencies as $z$ increases; binaries
at higher $z$ will be caught closer to coalescence and will therefore
show a lower residual eccentricity.

\begin{figure}
  \hspace{-0.9cm}
  \includegraphics[width=1.15\linewidth]{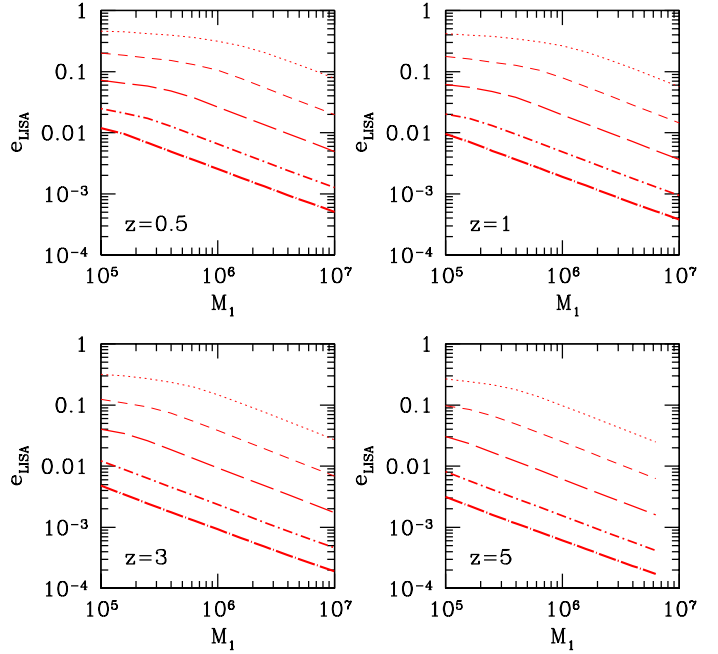}
  \caption{Residual eccentricity $e_{\rm LISA}$ as a function of $M_1$, for
different mass ratios. Each panel refers to BH binaries at different
redshifts as labelled in the figure. In each panel, from bottom to top,
curves are for ${\rm log}q=0,-0.5,-1,-1.5,-2$. }
 \label{fig:eccsurvival}
\end{figure}

\begin{figure}
  \hspace{-0.9cm}
  \includegraphics[width=1.15\linewidth]{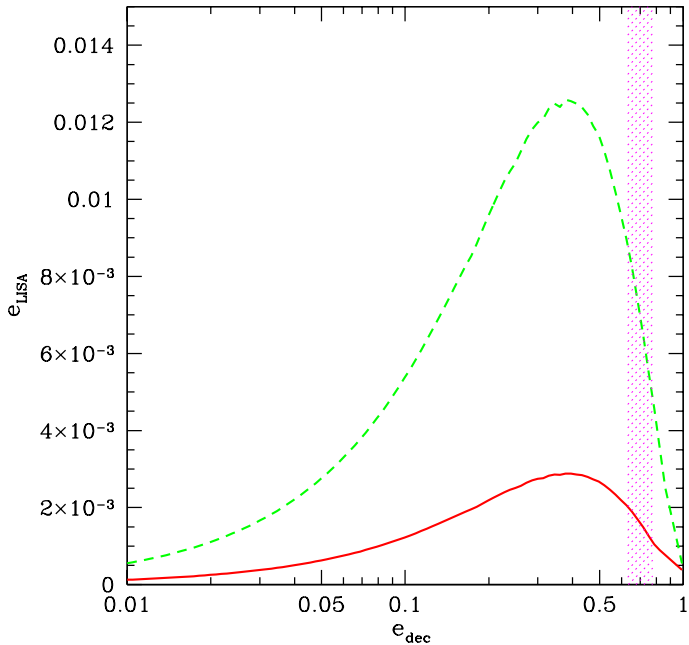}
  \caption{Residual eccentricity $e_{\rm LISA}$ as a function of $e_{\rm dec}.$
Red--solid curve refers to $q=1/3$, green--dashed curve to $q=0.1$. In the figure
the mass of the primary BH black hole is $M_1=2.6\times 10^6\,{\rm M_\odot}$ and the redshift of
the binary is $z=1$. The shaded vertical stripe brackets the limiting
eccentricity interval found in our simulations.}
 \label{fig:eccsurvival_e0}
\end{figure}

The predicted values of $e_{\rm LISA}$, as a function of $M_1$ for different $q$
and $z$,  are shown in Fig.~\ref{fig:eccsurvival}.
Not surprisingly, the residual eccentricity is
larger for lighter binaries (i.e., for lighter $M_1$)
and smaller mass ratios $q$. This is simply a consequence of the
scaling with $M$ and $q_{\rm s}$ of the frequency at decoupling,
$f_{\rm dec}$, and can be easily understood analytically as follows.
By coupling the orbital decay rate to the eccentricity decay
rate in the quadrupole approximation (sufficient for a scaling
argument, \cite{Peters:1963ux}),we get
\begin{equation}
\frac{f_{\rm r}}{f_o} = \left\{ \frac{1-e_o^2}{1-e^2} \left(\frac{e}{e_o}\right)^{\frac{12}{19}} \left[\frac{1+\frac{121}{304}e^2}{1+\frac{121}{304}e_o^2}\right]^{\frac{870}{2299}} \right\}^{-3/2},\label{eq:f_e_relation}
\end{equation}
where $f_{\rm r}=2f_{\rm K}$ is the frequency of the fundamental GW
harmonic (in the rest-frame of the source) inferred from
Kepler's law  $a^3=GM/(2\pi f_{\rm K})^2$. Eq.~(\ref{eq:f_e_relation})
allows us to compute $e$ at any given frequency $f_{\rm r}$, once
$e_o$ and $f_o$ are provided. In  our case  $e_o=e_{\rm dec}\sim 0.6$,
and $f_o=f_{\rm dec}(a_{\rm dec}).$ If we set the value $f_{\rm LISA}=10^{-4}$ Hz
as final
frequency,
Eq.~\ref{eq:f_e_relation}, in the limit of small final $e,$ gives
\begin{equation}
e_{\rm LISA}\propto f_{\rm dec}^{19/18}.
\label{eq:f_e_relation2}
\end{equation}
The identity $t_{\rm m}=t_{\rm GW}$ requires $a_{\rm dec}\propto M^{23/29}q_{\rm s}^{22/29}$.
Coupling this result to Kepler's law (i.e., $a^3\propto M f_{\rm r}^{-2}$), we get
$f_{\rm dec}\propto M^{-20/29} q_{\rm s}^{-33/29}$. Finally, using Eq.~\ref{eq:f_e_relation2} we obtain
\begin{equation} e_{\rm LISA}\propto M^{-0.73}q_{\rm s}^{-1.2},
\end{equation}
which is basically the $M$ and $q$ dependence observed in
Fig.~\ref{fig:eccsurvival}.
Fig.~\ref{fig:eccsurvival_e0} shows how this result depends on the
binary eccentricity at decoupling.  We see two interesting things:
firstly, there is a maximum $e_{\rm LISA}$ at $e_{\rm dec}\approx
0.4$ (i.e. $e_{\rm LISA}$ is not a monotonic function of $e_{\rm
  dec}$); secondly, as long as $0.1<e_{\rm dec}<0.7$, $e_{\rm LISA}$
changes only within a factor of $\approx 2$. This is a consequence of
the $t_{\rm GW}$ dependence on $e$. The higher $e$, the faster the
GW driven evolution, and  the larger is $a_{\rm dec}$. Even though
$e_{\rm dec}$ is larger, the binary has much more time to circularize
before entering the {\it LISA} band, showing a smaller residual eccentricity
$e_{\rm LISA}$.  We note that the exact value of $e_{\rm LISA}$
depends on the disc properties.  It is, however, interesting that a
small $e_{\rm LISA}$ can be associated both to a fairly circular
$e_{\rm dec}\approx 0.05$ binary or to a binary with $e_{\rm
  dec}>0.95$.

These results obviously depend on the assumed disc parameters. Both
a lower $\dot{m}$ and a lower $\alpha$ would increase $t_{\rm m}$, resulting
in a larger $a_{\rm dec}$ and, in turn, in a smaller $e_{\rm LISA}$. On the
other hand, if the BHs have large spins, the radiative efficiency $\epsilon$ may be up to a
factor of 3 larger, acting in the opposite direction. It is however worth
to keep in mind that $t_{\rm GW}\propto a^{4}$. A change of
a factor of 10 on $t_{\rm m}$ will therefore result in a change of
about $\sim1.8$ $a_{\rm dec}$, eventually influencing $e_{\rm LISA}$ only
by a factor of two. We can therefore consider our results
robust and only mildly dependent on the details of the disc.



\section{Conclusions}
\label{sec:conclusions}

In this paper, we explored the dynamics of sub-pc BH binaries
interacting with a circumbinary gaseous disc after they have
excavated a gap in the surface density distribution.  We ran
a sequence of numerical models that differ only in the initial binary eccentricity
$e_0$. Our aim was to study the evolution of the eccentricity in order
to answer the following question: does the eccentricity (which is
known to increase in initially circular binaries) continue to grow up
to $e\to 1$ so that BH binaries in such discs reach the GW domain on a
nearly zero angular momentum orbit, or does $e$ saturate, and if so, at
which value?

The key finding is that $e$ converges to a limiting value
$e_{\rm crit}.$
Binaries that start
with low eccentricities ($e_0< e_{\rm crit}$) increase $e$ up to
$e_{\rm crit}$, whereas binaries that start with high eccentricities
($e_0>e_{\rm crit}$) display the opposite behaviour, i.e. their
eccentricity declines with time approaching $e_{\rm crit}$.
Saturation rises due to the opposing action of the gravitational drag
experienced by the lighter, secondary BH in its motion near apo-apsis.
For low eccentricity orbits, the secondary BH excites a density
wake which lags behind the BH at apo-apsis,
causing its deceleration (and so a rise of $e$).  The opposite occurs
for a highly eccentric orbit: the
secondary moves more slowly than the disc (i.e. its angular frequency
is smaller than the angular frequency of the adjacent fluid elements)
and the density wake moves ahead of the BH path, causing a net
acceleration.
Using this simple analytical argument, the limiting eccentricity is
independent on the binary mass ratio, but is a function of the location
$\delta$ of the inner rim of the disc from the system center of mass.
For the range of values $2<\delta<2.5$, this argument predicts
$0.55<e_{\rm crit}<0.79$, consistent with our numerical findings.
The larger the gap size, the higher $e_{\rm
crit}$, the longer is the time scale on which this limit is
attained.  The expectation is that BH binaries, immersed in
circumbinary discs, maintain a large eccentricity throughout the
migration process. 
Althought in this study we have focused on BH binaries, the evolution of proto-stellar binaries occurs in a similar geometry \citep[e.g.,][]{Bate97,Artymowicz1994} and share much of the same physics.  Thus, our results are likely relevant for the interpretation of the observed distribution of binary star eccentricities \citep[e.g.,][]{Pourbaix04}.

The existence of a relatively large limiting eccentricity in a BH
binary, that emerges from the migration phase, has two important
observational consequences.  Firstly, the possibility of triggering
periodic inflows of gas onto the two BHs. This would enhance the
possibility of an electromagnetic identification of a sub-parsec BH
binary. Here we showed that periodicities occur on the dynamical time
related to the Keplerian motion of the binary (depending on the binary
parameters, from months to hundreds of years) and of the inner rim of
the circumbinary disc, together with the beat frequency between the
two. These features should be discernible in the power spectra of
active nuclei, and this issue will be explored in detail in a
forthcoming paper.
Secondly, a feasible GW signature of a BH binary, that evolved through
disc migration, is a detectable residual eccentricity at the time of
entrance in the {\it LISA} band.  In the case of our setup this
residual $e$ would amount to $e_{\rm LISA}\sim 2\times 10^{-3}$ for a
coalescing source at $z=1$, but can be as high as $e_{\rm LISA}>0.1$
for a lower mass, lower $q$ binary (with $M\sim 10^5\,\rm {M_\odot}$
and $q<0.1$) at the same redshift.  Thus, this study has an impact
both on searches of periodicities in the light curves of active BHs,
as well as on GW data stream analysis.


\section*{Acknowledgement}
We thank the anonymous referee for suggestions that greatly improve the presentation of our results.  We thank Pau Amaro Seoane, Luciano Rezzolla and Julian Krolik for useful discussions.
CR wishes to thank Nico Budewitz for support in all matters HPC. The computations were
performed on the {\it damiana} cluster of the AEI.  JC acknowledges support from
FONDAP Center for Astrophysics (15010003), FONDECYT (Iniciaci\'on 11100240) and VRI-PUC (Inicio 16/2010).

\bibliographystyle{mn2e}
\bibliography{biblio/aeireferences}

\bsp

\label{lastpage}

\end{document}